\documentclass[runningheads]{llncs}
\usepackage[misc]{ifsym}
\usepackage{xcolor}
\usepackage{epsfig}
\usepackage{graphicx}
\usepackage{amsmath}
\usepackage{amssymb}
\usepackage{svg}
\usepackage{amsmath}
\usepackage{booktabs}
\usepackage{bm}
\usepackage{bbm}
\usepackage{upgreek}
\makeatletter

\usepackage{graphicx}

\usepackage{multirow}
\usepackage{boxhandler}
\usepackage{tabularx}
\usepackage{adjustbox}
\usepackage{epstopdf}
\usepackage{wrapfig}
\usepackage{makecell}

\begin{document}

\title{Disguising Personal Identity Information in EEG Signals}
%
%

\author{ Shiya Liu \and Yue Yao \and Chaoyue Xing  \and Tom Gedeon$^{(\textrm{\Letter})}$}
\authorrunning{S. Liu et al.}
%
\institute{Research School of Computer Science\\ Australian National University \\
\email{\{shiya.liu, yue.yao, u6920870\}@anu.edu.au}, 
\email{tom@cs.anu.edu.au}
}
\tocauthor{Shiya Liu, Yue Yao, Chaoyue Xing, and Tom Gedeon}
\toctitle{Disguising Personal Identity Information in EEG Signals}

\maketitle              
\begin{abstract}
There is a need to protect the personal identity information in public EEG datasets. However, it is challenging to remove such information that has infinite classes (open set). We propose an approach to disguise the identity information in EEG signals with dummy identities, while preserving the key features. The dummy identities are obtained by applying grand average on EEG spectrums across the subjects within a group that have common attributes. The personal identity information in original EEGs are transformed into disguised ones with a CycleGAN-based EEG disguising model. With the constraints added to the model, the features of interest in EEG signals can be preserved. We evaluate the model by performing classification tasks on both the original and the disguised EEG and compare the results. For evaluation, we also experiment with ResNet classifiers, which perform well especially on the identity recognition task with an accuracy of 98.4\%. The results show that our EEG disguising model can hide about 90\% of personal identity information and can preserve most of the other key features. Our code is available at \url{https://github.com/ShiyaLiu/EEG-feature-filter-and-disguising}. 

\end{abstract}
\section{Introduction}
\label{cha:intro}
Electroencephalography (EEG) is one of the most common methods of measuring brain electrical activities \cite{ubeyli2009combined}. The signals contain complex information about the state of the human brain, which can be used to study brain functions, detect brain abnormalities and so on. Due to the uniqueness of the EEG signal from every individual, it can be a reliable source of biometric identification \cite{campisi2014brain,palaniappan2007biometrics}; on the other hand, such personal information could be used for malicious purposes if not well protected. With increasing interest in and demand for research on EEG signals, there has been many EEG databases disclosed to the public. Therefore, there is a need to protect from exposure the personal identity information in the EEG signals of the experiment subjects, who will kindly contribute to such public EEG datasets in the future.

Our key contribution is the creation of labelled EEG with dummy identities mimicking EEG signals from groups that have common features, but with averaged identity information. The EEG data with dummy identities helps disguise the identity information in the original EEG by composing a training set in the target domain, to which the original EEG is transformed. The experiment results suggest that the dummy identities effectively hide real identities in EEG, while our approach keeps the features of interest.

\section{Related Works}
\label{cha:background}
Yao et al. proposed a feature filter to protect certain diseases from exposure  \cite{yao2020information}. There, the EEG signals with disease information are transferred to healthy EEG signals without such disease, so that the disease information can be effectively removed. The limitation of that approach is that it can only deal with the features from a limited number of categories (closed set), such that the EEG signals can be transferred from one class to another. For information that has potentially infinite classes (open set), such as personal identity, it is not feasible to find a class of EEG signals that has no identity information.

Instead of removing the information related to biometrics in EEG, this paper proposes an approach of disguising the identity information using a dummy identity, such that the EEG signal cannot be used for person recognition.

\section{Methods}
\label{cha:methodology}
\subsection{UCI EEG Dataset}
The time-series data consists of EEG signals from two groups of subjects – alcoholic individuals and controls \cite{begleiter1999eeg}. The dataset has three attributes: alcoholism, stimulus condition and subject identity. Each EEG signal contains $64 \text{ channels} \times 256 \text{ samples}$ of data. The data within each subject is split into a training set (70\%), a testing set (20\%) and a validation set (10\%), using the same within-subject train-test-validation splitting method as \cite{li2017targeting,yao2020information}.

\subsection{EEG Data Pre-processing}
The raw time-domain EEG data is challenging to analyse due to its high dimensionality. We can extract prominent features from the frequency bands in EEG spectrum \cite{mcgrogan1999neural,subasi2005automatic} for dimension reduction. We convert the EEG signals into the frequency domain using Fast Fourier Transform (FFT), which can capture most characteristics from stationary signals; and the short-term EEG signals in this work can be regarded as stationary \cite{tcheslavski2012alcoholism}.

\begin{figure}[t]
	\centering
		\includegraphics[width=0.6\linewidth]{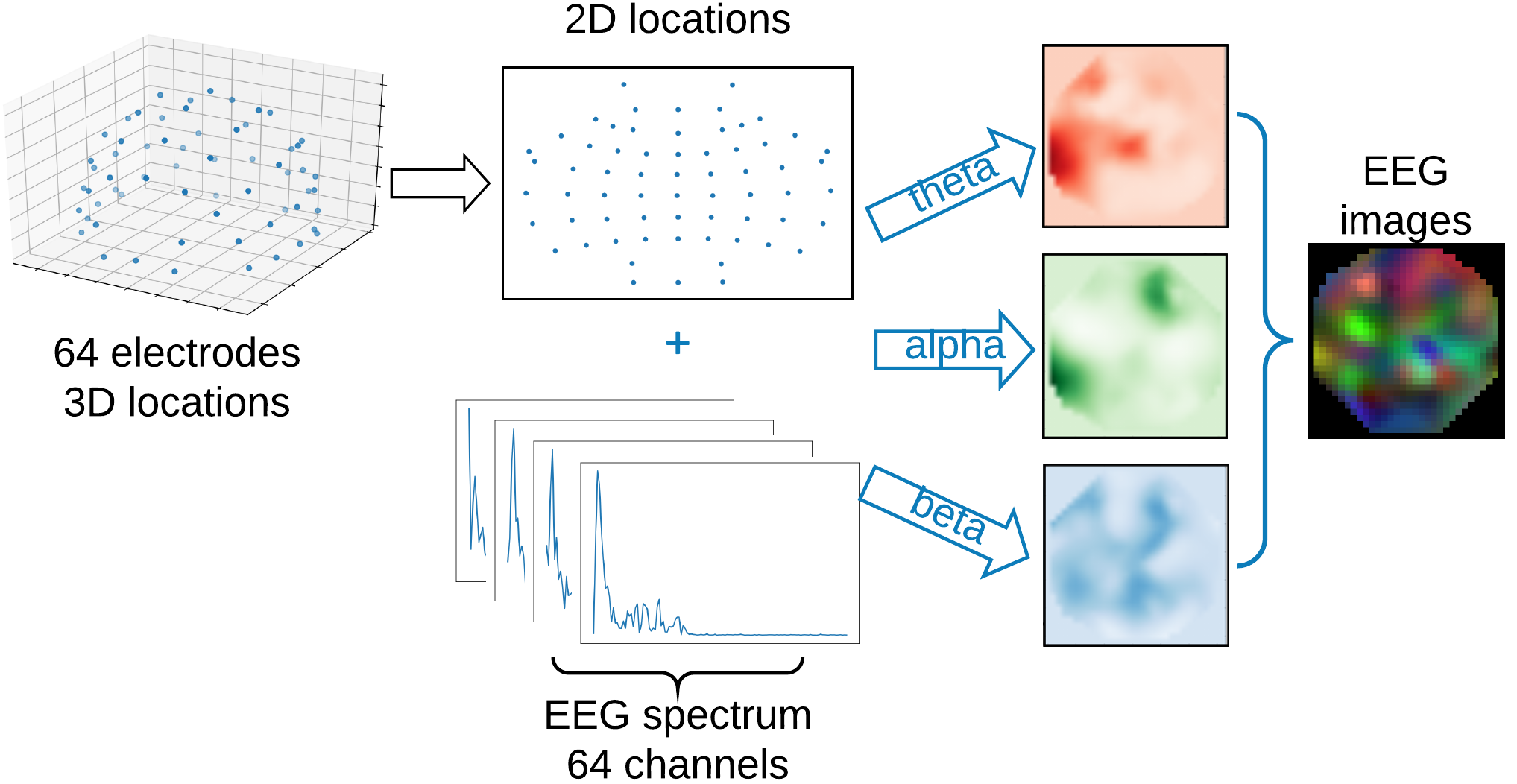}
		\caption[Convert EEG spectrum into RGB EEG images] {Convert EEG spectrum into RGB EEG images \cite{yao2020information}. Features from three frequency bands are extracted and used as red, green and blue channels of an image.}
        \label{figure:EEG2img}
\end{figure}
To extract lower-level spectrum features, we adopt an approach proposed by Bashivan et al., which captures both spectral features and spatial information and represents them in images \cite{bashivan2015learning}. As in Fig. \ref{figure:EEG2img}, with the spectrum features of the 64 channels and their corresponding electrode locations projected on a 2D plane, a feature matrix is calculated for each frequency band. we obtain color images by merging the feature matrix from three key frequency bands. As the EEG signals are collected from subjects as evoked by visual stimuli, we 
select the mid range frequency bands, i.e. bands $\theta$ (4-8 Hz), $\alpha$ (8-13 Hz) and $\beta$ (13-30 Hz), which are less noisy and captures the most related information \cite{musha1997feature}.

\subsection{Disguised EEG Images Generation}
\subsubsection{Obtaining Labelled EEG Images with Dummy Identities}
\label{sec:dummyEEG}
In neuroscience, grand averages of EEG signals across subjects is a common technique when statistically analysing EEG patterns in certain conditions \cite{delorme2015grand,polich1997eeg,marshall2007infant}. The waveform of the grand mean of EEG signals can be regarded as a representative of a group and used to study their significant characteristics. A similar technique is also used in computer vision (CV) to investigate facial characteristics. E.g., Burt and Perrett generated composite face images by averaging the faces from different age groups, and those average faces can still be correctly recognized as faces in the corresponding age range \cite{burt1995perception}. 

We adopt this technique to generate average EEG images, which hold the characteristics of the common features of EEG signals in a certain group while having average biometric information. The averaged biometric information can be regarded as dummy identities. As the EEG dataset has 2 classes for the alcoholism attribute and 5 classes for stimulus condition attribute, we split the dataset into 10 groups corresponding to the 10 combinations of attributes. Then we take the grand average power of the EEG spectrums across each of several subjects within each group. The average EEG spectrums are converted into EEG images (see section 3.2). This process is applied to the training dataset and the average EEG images are labelled as they are obtained within each group.

\subsubsection{From Original to Disguised}
We train a CycleGAN model \cite{zhu2017unpaired} to generate the disguised EEG images that have the features which are consistent with the corresponding original images, but with dummy identity information instead of real identity information. For the training datasets, the source-domain data are the real EEG images and the target-domain data are the average EEG images with dummy identity obtained in section \ref{sec:dummyEEG}.  

The EEG images used in the model are labelled and we want them to be correctly classified to the original labels after translation to the target domain with dummy identities. Therefore, we need further constraints on our model to minimize the information loss in the most interesting features. We apply task and semantic loss in our model, as proposed by Hoffman et al. in their Cycle-consistent Adversarial Domain Adaptation (CyCADA) model \cite{hoffman2017cycada,yao2020simulating}. 

\begin{figure}[t]
	\centering
		\includegraphics[width=0.8\linewidth]{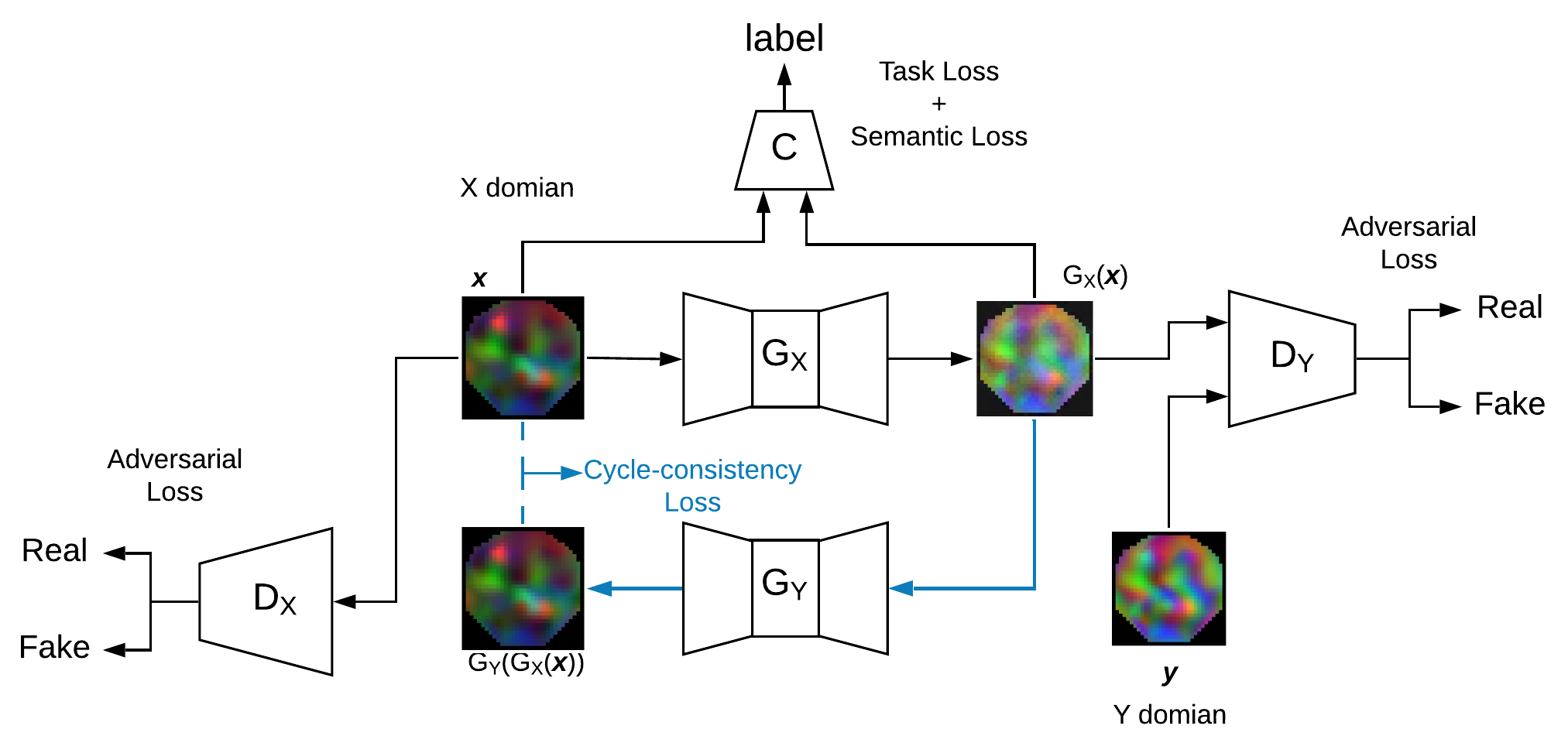}
		\caption[The architecture of the image disguise model] {The architecture of the image disguising model with task loss and semantic loss. $G_X$ and $G_Y$ denote the generators with real input images from X and Y domain respectively; $D_X$ and $D_Y$ denote the discriminator that distinguishes whether images are from X and Y domains respectively; $\bm{x}$ and $\bm{y}$ denote input images from X domain and Y domain. $G_Y(G_X(\bm{x}))$ is a reconstructed image in X domain, which should be close to the corresponding real images $\bm{x}$; $C$ denotes the classifier that is used to add a further constraint to the model. For clearer illustration, only one direction of the cycle ($\bm{x}\rightarrow G_X(\bm{x})\rightarrow G_Y(G_X(\bm{x}))$) is shown.}
        \label{figure:cycada}
\end{figure}
We add an additional classifier C to our own CycleGAN-based model (Fig. \ref{figure:cycada}) \cite{hoffman2017cycada}. First, the classifier learns to make predictions on the labels of EEG images by minimizing the task loss during training. After the classifier is sufficiently trained to have relatively low task loss (\textless{1.0}), the model can take the semantic consistency into account, which means the label of a disguised EEG $G_X(\bm{x})$ predicted by the classifier should be consistent with the predicted label of the original EEG $\bm{x}$, and similar to the other generator $G_Y$. The generators should learn to minimize the semantic loss during training.

\subsection{Classification Tasks}
To evaluate our model, we need to validate whether the personal identity information has been disguised such that it fails the person recognition task; also, we need to validate whether the information of interest is preserved, which means a disguised EEG image should be correctly classified to its original label. Therefore, we need to train classification models to predict the labels of the EEG data with respect to original subject identities, alcoholism and stimulus conditions. 

\subsubsection{ResNet Classifier}
We could use a deep CNN model to extract high-level features from EEG images, which contains the low-level features we obtained from the EEG spectrum. Although deeper networks may give better results with higher-level features extracted, they also bring difficulties to optimization during the training and as a result degrade the performance in practice; this problem can be addressed by deep residual learning \cite{he2016deep}. Thus, we implement ResNet models for the classifications which are essential to the evaluation step. To explore the impact of depth of ResNet models on our classification tasks, we will experiment with 18-layer, 34-layer and 50-layer ResNet models.

\subsubsection{Joint Training} Since the average EEG images obtained in section \ref{sec:dummyEEG} are labelled data, we make use of them for joint training by combining them with the training dataset when training the classification model for alcoholism detection and stimulus classification tasks. This can add randomness and diversity to the training dataset thus improving the generalization of the classification model \cite{krizhevsky2012imagenet}.

\section{Experiment Results and Discussion}
\label{cha:result}
\subsection{Evaluation Criteria}
We aim for the disguised EEG images to fail the person recognition task while not sacrificing the performance on the alcoholism detection task and the stimulus classification task, and $accuracy = \frac{\text{TP + TN}}{\text{TP + FP + TN + FN}}$ is an important criterion when measuring those results. When classifying the subject identities of the disguised EEG, the accuracy should drop significantly compared with the results for original EEG images using the same classifier, while the accuracy of detecting alcoholism and classifying stimulus conditions of the disguised EEG images should be at the same level as that of the original EEG images.

Both the test dataset and validation dataset are imbalanced in the alcoholism label, so it is biased if accuracy is the only evaluation criterion. Thus, we also use $sensitivity = \frac{\text{TP}}{\text{TP + FN}}$ and $specificity = \frac{\text{TN}}{\text{TN + FP}}$ as additional evaluation criteria for the alcoholism detection task.

\subsection{Performance of the Classification Models}
To evaluate the performance of our EEG disguising model, we need to first train a classification model. We experiment with ResNet-based classification models with different numbers of layers. For comparison, we also trained an autoencoder-based classification model from Yao et al. \cite{yao2020information}.
\begin{table}[t]
\centering
\caption{The testing results of classification models. Our Resnet based model outperfom autoencoder based model by 18.87\% in terms of Identity accuracy. }
\label{results_cls}
\begin{tabular}{@{}|l !{\vrule width 1pt} c|c|c !{\vrule width 1pt} c !{\vrule width 1pt} c|@{}} 
\hline
            \multirow{2}{*}{} & \multicolumn{3}{c!{\vrule width 1pt}}{Alcoholism (\%)} & \multicolumn{1}{l!{\vrule width 1pt}}{Stimulus (\%)} & \multicolumn{1}{l|}{Identity (\%)} \\ \cline{2-6} 
            & Acc.    & Sens.   & Spec.   & Acc.    & Acc. \\ \Xhline{1pt}
Yao et al. & 87.84   & 88.16   & 84.42   & 51.67   & 79.53 \\
ResNet-18   & \textbf{92.49}   & \textbf{91.13}   & \textbf{93.85}   & 61.98   & \textbf{98.40}  \\
ResNet-34   & 91.88   & 90.78   & 92.99   & \textbf{62.69}   & 97.26   \\
ResNet-50   & 87.11   & 83.76   & 90.49   & 60.91   & 90.49 \\   
\hline
\end{tabular}
\end{table}

As shown in Table \ref{results_cls},  the ResNet-18 classification model performs well on alcoholism and personal identity recognition, with an accuracy of 92.5\% and 98.4\% respectively. In addition, the model achieves 91.1\% sensitivity and 93.9\% specificity in the alcoholism detection task, which is a balanced result. The testing results indicate that the ResNet-18 model has good generalization when classifying unseen samples. The model predicts the stimulus condition with an accuracy of 62.0\%, which is much higher than chance (20\%) although it may not be optimal. The ResNet-34 classifier achieves similar results (slightly lower) compared to the ResNet-18 model and only the accuracy of stimulus condition prediction task (62.7\%) is slightly higher, which indicates that the deeper network may contribute slightly to this task. However, the performance degrades when we explore deeper networks using a ResNet-50 classifier (Table \ref{results_cls}).

The results in Table \ref{results_cls} show that the ResNet models outperform Yao et al.'s autoencoder-based model in general. With deeper network and more free parameters, the ResNet models are able to extract enough higher-level features for the classification tasks, especially the personal identity recognition, on which the ResNet models achieve the most improvement. Although the ResNet models benefit from deeper networks, the model with the most layers does not achieve the best results. The performance of the ResNet-50 model drops significantly. It indicates that the model is overly deep and has too many parameters for these EEG- related classification tasks, which causes overfitting.

\subsection{Evaluation on the EEG Disguising Model}
We use trained ResNet-18 models for the personal identity, alcoholism and stimulus condition classification tasks on both the original validation dataset and the corresponding disguised EEGs. In this experiment, we assume that the alcoholism information is of more interest; so during the training, the semantic loss and task loss come from the classifier’s prediction on the alcoholism label. 
\begin{figure}[t]
	\centering
		\includegraphics[width=0.8\linewidth]{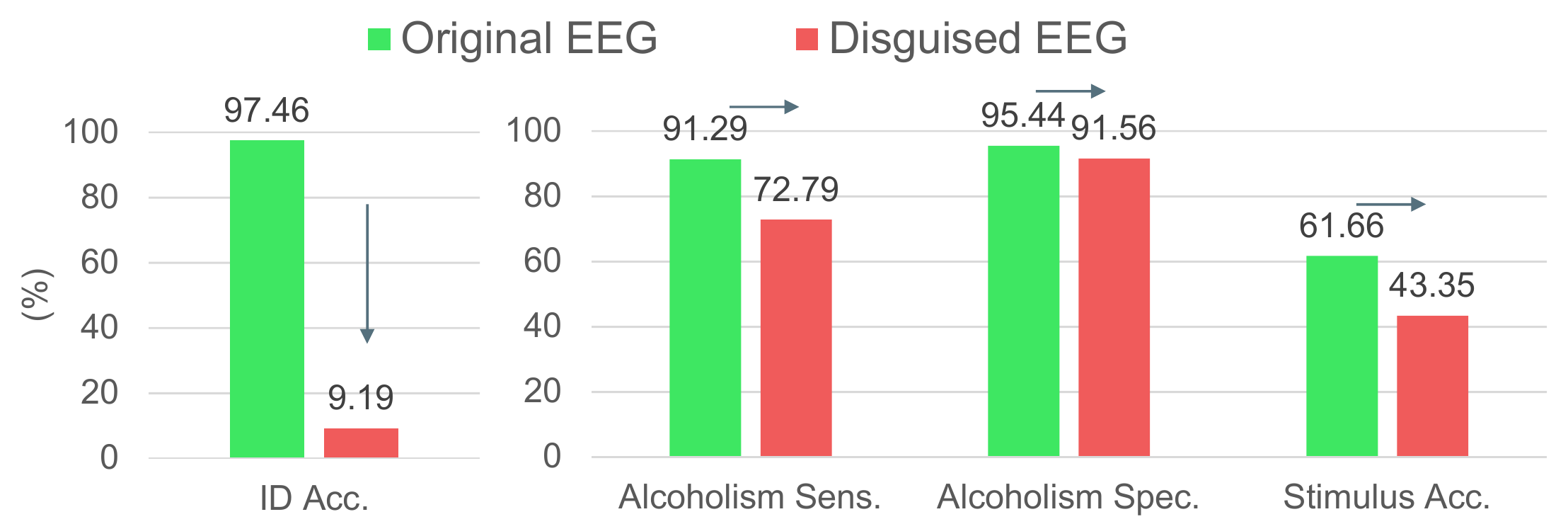}
		\caption[Classification results: original EEG vs. disguised EEG] {Classification results: original EEG vs. disguised EEG. The disguised images generated by our method hides about 90\% of personal identity information and can preserve most of the key features used for alcoholism and stimulus prediction. }
        \label{figure:cls_bar_EEG}
\end{figure}

As shown in Fig. \ref{figure:cls_bar_EEG},  when classifying the personal identity of the original EEG images, the accuracy is 97.5\%; after those are disguised, the accuracy drops dramatically to 9.2\%. It indicaes that the personal identity information in the original EEG is successfully hidden by our EEG disguising model and cannot be recognized by the classifier, which performs well on the original EEG data. For alcoholism detection, the classifier performs well on the original EEG images, achieves 91.3\% sensitivity and 95.4\% specificity; when using the same model to classify the disguised EEG images, the specificity only drops to 91.6\%, while the sensitivity drops to 72.8\%.  For the stimulus condition classification task, although the accuracy drops from 61.7\% to 43.4\% when the task is performed on the disguised EEG, it is still much higher than chance (20\%). 

Although the decrease in the results of alcoholism and stimulus classification tasks show that there is some information loss in terms of the corresponding features, we do not see a sharp fall comparable with the results of the identity recognition task. It indicates that most of the alcoholism and stimulus features are preserved such that the classification model can still make similar predictions.

\subsection{Ablation Study on the EEG Disguising Model}
\begin{table}[t]
\centering
\caption[Classification results (\%): disguised EEG with different semantic constraints]{Classification results (\%): disguised EEG with different semantic constraints. (a) Results with original EEG. (b) The 1st row (baseline) shows results without semantic constraints; the 2nd row (+ Alc.) shows results with the semantic constraint on the alcoholism feature; the 3rd row (+ Sti.) shows results with the semantic constraint on the stimulus feature; the 4th row (+ Alc. \& Sti.) shows results with the semantic constraint on both of the alcoholism and stimulus feature.}
\label{results_disguised_eeg}
\begin{tabular}{|c !{\vrule width 1pt} c|c|c|c|}
\multicolumn{5}{l}{(a) Original EEG} \\ \hline
Original EEG        & ID Acc.   & Alcoholism Sens.  & Alcoholism Spec.  & Stimulus Acc. \\ \Xhline{1pt}
                    & 97.46     & 91.29             & 95.44             & 61.66  \\ \hline
\multicolumn{5}{l}{} \\
\multicolumn{5}{l}{(b) Disguised EEG with Different Constraints} \\ \hline
Constraint on       & ID Acc. $\downarrow$ & Alcoholism Sens. $\uparrow$  & Alcoholism Spec.$\uparrow$  & Stimulus Acc.$\uparrow$ \\ \Xhline{1pt}
Baseline    & 0.48      & 65.17             & 35.41             & 37.79  \\ \hline
+ Alc. & \textbf{9.19}   & 72.79     & \textbf{91.56}             & 43.35  \\ 
+ Sti. & 21.19    & 78.91     & 62.38             & \textbf{52.61}  \\ 
+ Alc. \& Sti.   & 48.97      & \textbf{93.47}             & 64.59             & 50.41  \\ \hline
\end{tabular}
\end{table}
\subsubsection{Ablation: Semantic Constraints on Stimulus Condition Feature} We also use the classifier in our EEG disguising model to impose a semantic constraint associated with the stimulus condition feature on the disguised EEG. Table \ref{results_disguised_eeg} shows that the stimulus classification task on the disguised EEG achieves 52.6\% accuracy, which is significantly higher than the counterpart result of using the classifier for alcoholism feature (43.4\%). It implies that the classifier can effectively preserve more information of interest. Also, the results show that the performance of alcoholism detection task degrades, as the loss of alcoholism information increases without the constraint on it. Further, we find that the accuracy of personal identity task on disguised EEG also increases to 21.2\%, which implies that the constraint may be too strong for the EEG disguising model so that more personal identity information is kept to reduce the semantic loss. A possible reason could be the classifier relies more on the identity information to predict the stimulus condition rather than the actual stimulus features. 

\subsubsection{Ablation: Semantic Constraints on both Alcoholism and Stimulus} We explore more restricted semantic constraints on the model by using the classifier to predict both the alcoholism and stimulus labels when calculating semantic loss. The results in Table \ref{results_disguised_eeg} show the accuracy of the stimulus condition classification task (50.4\%) and the sensitivity of the alcoholism task (93.5\%) are higher, compared to the results of the models without semantic constraints on the stimulus or alcoholism features, respectively. However, the specificity of the alcoholism detection task is low, one possible reason is that the classifier may not be able to well handle the multi-label classification problem. The accuracy of personal identity increases as the constraint has become too strong.  
\subsubsection{Ablation: No Additional Semantic Constraints} We experiment with the model without the additional classifier that places semantic constraints on the disguised EEG. The results (Table \ref{results_disguised_eeg}) shows that the performance degrades on both the alcoholism and the stimulus condition classification task with the disguised EEG. It demonstrates that the semantic constraints are critical in our model to preserve the information of interests. Correspondingly, the accuracy of personal identity recognition (0.48\%) drops significantly compared with the results of the model with the constraints. Without constraints, more identity information in the original EEG is disguised.  

\section{Conclusion and Future Work}
Our EEG disguising model can be used to protect personal privacy, by hiding the personal identity information in EEG signals. The results demonstrate that the model is able to disguise 90\% of the personal identity information in EEG signals with dummy identities while preserving most of the key information. In addition, we experiment with ResNet classifiers that can be used to perform different EEG classification tasks. From the results, we find that ResNet models are suitable for complex EEG signals, especially for the personal identity recognition, which requires more parameters and higher-level features to solve.

The information loss during the EEG disguising process should be improved, we may use a more complex classifier as the constraint. Also, we will validate our model on other EEG datasets to determine experimentally how well the model works in general. In addition, we will explore other techniques to improve the performance of the stimulus condition classification task. As an extension, we will also explore the possibility of decoding EEG signals at the feature level, one hypothetical approach could be gradually filtering out different key features in an EEG signal to see whether we can decode the EEG signal eventually. 

%
%
%

\end{document}